\documentclass[a4paper,11pt]{article}
\pdfoutput=1 

\usepackage{jheppub} 

\usepackage[vcentermath]{youngtab}
\usepackage{braket}
\usepackage{array}
\newcolumntype{C}[1]{>{\centering\arraybackslash$}p{#1}<{$}}

\newcommand{\yl}[2]{\Ket{\left[#1\right]#2}}

\newcommand{\s}{$S$}
\newcommand{\gtilde}{$\tilde{g}$}

\begin{document} 

\title{\boldmath Wave function of the sexaquark or compact H-dibaryon}

\author[a,1]{Glennys R. Farrar}
\author[b,2]{and Nico Wintergerst}

\affiliation[a]{Center for Cosmology and Particle Physics, New York University, New York, NY, USA}
\affiliation[b]{The Niels Bohr Institute, Blegdamsvej 17, DK-2100 Copenhagen \O, Denmark}

\emailAdd{gf25@nyu.edu}
\emailAdd{nwintergerst@gmail.com}

\abstract{We derive and explicitly display the  internal wave function of a color-, flavor- and spin- singlet dibaryon composed of $uuddss$ quarks in a spatially symmetric state, in the approximation of exact SU(3) flavor symmetry.  
This wavefunction shows that the often-used superposition of $\Lambda \Lambda$, $N \Xi$ and $\Sigma \Sigma$ baryons, relevant for di-baryon molecules, accounts for only 1/5 of a spatially symmetric six-quark color-flavor-spin-singlet state, with the remaining 4/5 consisting of products of color-octet baryons.  Using the correct wavefunction has important implications for calculations of the mass of the lowest-lying flavor-singlet dibaryon, as we illustrate using the Cornell potential.  We also provide a compact representation of the state in terms of creation operators, and comment on the impact of not using the optimal operator in lattice studies of this system.}

\maketitle
\flushbottom

Data from ALICE~\cite{ALICE_LamLam19}, and the latest lattice QCD calculations~\cite{SasakiHALQCD20,greenMainzH21}, indicate that the color-, flavor- and spin- singlet H-dibaryon suggested by Jaffe~\cite{jaffe:H}, is a near-threshold resonance or very loose bound state of two $\Lambda$'s.  In addition to this spatially extended molecular configuration, the same $uuddss$ quarks may form a deeply bound compact state which would naturally be very long lived~\cite{fzNuc03,fS17} and could be a dark matter candidate~\cite{fudsDM18}. This compact state was designated \s\ or sexaquark to avoid confusion with the H;  for an overview of constraints and why it would have escaped detection see~\cite{fS22,xf23a,fw23a}.   

Deciding whether a deeply bound state of $uuddss$ quarks exists is an important task for lattice QCD, because detecting it in the laboratory is surprisingly challenging~\cite{fS22}.  An important element of a robust lattice QCD calculation is using a source which has a strong overlap with the true state.  Even with an optimal overlap, the calculation is difficult due to the rapid growth of noise with the number of light quarks~\cite{LepageTASI} and the difficulty of separating out multiple contributing energy levels~\cite{greenMainzH21}.

The wave function for a hadron can be written
\begin{equation}
   \Psi = \Psi_{cfs} \Psi(r_1, ... r_6)~, 
\end{equation}
where the spatial wavefunction $\Psi(r_1, ... r_6)$ is spatially symmetric for deeply bound states, with all quarks occupying the same wavefunction.  (We do not address the spatial wavefunction here.)  For a spatially-symmetric wavefunction, Fermi statistics requires that the internal wavefunction must be fully antisymmetric under interchange of any pair of quarks.

In this paper we derive and present in an explicit form, the wavefunction $\Psi_{cfs}$ for $uuddss$ simultaneously in a singlet of color, flavor and spin.  We work in the approximation of equal mass light quarks, so $m_u = m_d = m_s$.  This is a well-motivated first approximation since the current quark masses are small compared to the QCD scale.\footnote{Park, Park and Lee~\cite{pplSwfn16} provide an explicit representation of this wavefunction; our considerably more compact form is a more tractable complement.  (We thank A. Gal for bringing~\cite{pplSwfn16} to our attention, after v1 of our paper was posted to the arxiv.)  Ref.~\cite{pplSwfn16} also gives the wavefunction when flavor SU(3) is broken to isospin.  The first discussion of the effect of SU(3) flavor breaking, taking account of $m_s > m_u,\, m_d$ on the general structure of the spectrum but not constructing the wavefunction was given by Rosner~\cite{rosnerH86}.  Reference~\cite{MackenzieThacker85} gives a very early lattice QCD estimate of the H-dibaryon mass.  They note that obtaining the color-, flavor- and spin- wavefunction, which using their construction is too ``huge" to be given explicitly, was the computationally most challenging part of the project: ``The calculation of the H two-point function from the quark propagators performed in this way required two weeks of VAX 11/780 central-processing-unit time, compared with a few days for all the rest of the spectrum and coupling-constant analysis combined." }   
Using this wavefunction, we show that when six quarks are in a color-, flavor- and spin-singlet, only $\sqrt{1/5}$ of the amplitude is a product of color singlet baryons -- the other  $\sqrt{4/5}$ is a sum of products of color octet baryons. We illustrate, using the Cornell potential, that the presence of the color octet-octet component in a spatially symmetric flavor-spin singlet dibaryon significantly increases the strength of the binding. 

We re-express the explicit wavefunction compactly as entangled products of quark creation operators acting on the vacuum.  The operator version of the wavefunction shows that a specific entanglement is required to enforce the correct flavor-spin representation. Use of the creation operator we provide, or its equivalent, should thus improve the ability of lattice QCD to probe or exclude the existence of a deeply bound flavor-spin-singlet dibaryon without pollution from extraneous states in other flavor-spin representations. 

\section{How to construct the wavefunction}

Being a colorless, flavor-neutral spin-0 state, the S belongs to the $(1,1,1)$ representation of $SU(3)_C \times SU(3)_F \times SU(2)_S$, where subscripts denote color, flavor and spin, respectively. In a spatial $s$-wave state, in order to accommodate the Pauli principle, the S lives in the totally antisymmetric $6$-quark representation of $SU(18)$.  In Young tableau notation
\begin{equation}
\label{eq:tas}
\yng(1,1,1,1,1,1) = 18564 \, .
\end{equation}
The Young tableau notation has one box for each constituent quark -- here 6, because there are 6 quarks -- and the permutation symmetry of the state is characterized by the arrangements of the boxes into rows and columns.  Horizontal boxes are symmetrized and vertical boxes are anti-symmetrized.  The reader is referred to a text such as~\cite{hammermesh}, for additional background.

\subsection{Notation}
Before going into the details of the construction, let us briefly introduce our notation. Irreducible representations (irreps) will be referred to by their labels under the permutation group $S_6$: $\ket{[\mu]m}$. Here $\mu$ is a set of integers corresponding to the row lengths in the Young diagram and $m$ is the ordinal, obtained by ordering the tableaux according to their Yamanouchi labels (see, e.g.,~\cite{hammermesh}). The latter are obtained by filling an $n$-box diagram with the integers $1, ..., n$ according to specified rules and considering the rows of the numbers in decreasing order. The ordinal is obtained by ordering the Yamanouchi labels by their numerical value. For example,
\begin{equation}
\label{eq:yamlab}
\young(123,456) = (222111) = \yl{33}{1},\,\,\young(13,25,46) = (323121) = \yl{222}{4}~.
\end{equation}
As a representation of $S_6$, \eqref{eq:tas} is one-dimensional. We may thus omit the ordinal and refer to it as $\Ket{\left[1^6\right]}$.

\subsection{Wavefunction}
\label{sec:Young}
Since we are considering only $u,\, d, \, s$ quarks, there are three colors and three flavors and the maximum number of boxes in any column of the color and flavor Young tableaux is three, and similarly for spin with just two degrees of freedom the maximum height is two.
Diagrammatically, we are searching for the $\Ket{\left[1^6\right]}$ contained in the product of the $SU(3)_C$, $SU(3)_F$, $SU(2)_S$ singlets
\begin{equation}
\yng(1,1,1,1,1,1) \subset \yng(2,2,2) \times \yng(2,2,2) \times \yng(3,3)~.
\end{equation}
As a starting point, we note that a totally antisymmetric representation is only contained in a product of conjugate representations -- those whose Young tableaux are rotated by 90$^\circ$.  For instance 
\begin{equation}
\label{eq:C-FS}
\yng(1,1,1,1,1,1) \subset \yng(2,2,2) \times \yng(3,3) ~.
\end{equation}
It should also be noted that conjugate representations have a common dimension, where the dimension is the number of inequivalent realizations of the exchange symmetry.  The $\left[2^3\right]$ and the $[33]$ are five-dimensional.

Without loss of generality, let us first split the relevant groups into $SU(3)_C$ and $SU(6)_{FS}$. Then we will further split $SU(6)_{FS}$ to obtain the desired product of $SU(3)_C$, $SU(3)_F$, $SU(2)_S$ singlets.
Following the above statement and Eq.~\eqref{eq:C-FS}, we can write
\begin{equation}
\label{eq:cgc_singlet}
\Ket{\left[1^6\right]} = \sum_i^5 c_i \yl{2^3}{i}_C \yl{33}{6-i}_{FS}\,,
\end{equation}
where the $c_i$ are the Clebsch-Gordan coefficients. We have here used the fact that both the $\left[2^3\right]$ and the $[33]$ are five-dimensional.

We now decompose the $\yl{33}{k}_{FS}$ into a sum of products of irreps of $SU(3)_F$ and $SU(2)_S$.
\begin{equation}
\yl{33}{k}_{FS} = \sum_{l,m} d^k_{lm} \yl{2^3}{l}_F \yl{33}{m}_S \,,
\end{equation}
which leads to the following expression for the S wavefunction
\begin{equation}
\label{eq:hwave}
\Ket{S} \equiv \Ket{\left[1^6\right]} = \sum_{i,j,k} c_i\, d^{6-i}_{j,k} \,\yl{2^3}{i}_C \yl{2^3}{j}_F \yl{33}{k}_S\,.
\end{equation}
The nontrivial information is now contained in the Clebsch-Gordan coefficients $c$ and $d$.

\subsection{Clebsch-Gordan coefficients}
The eigenfunction method provides a simple and efficient way to calculate the Clebsch-Gordan coefficients (see~\cite{chen_review, chen_book} for reviews). For $n$-quark states, the method simultaneously diagonalizes the chain of operators 
\begin{equation}
{\cal C}(f) = \sum_{i < j}^f{\cal P}_{ij}\,,\hspace{20pt} f = 2,...,n ~,
\end{equation}
where ${\cal P}_{ij}$ permutes the $i$th and $j$th quarks.
The Clebsch-Gordan coefficients fulfill the eigenvalue equations (which can be derived using the usual properties of Young diagrams under permutations~\cite{chen_review, chen_book} )
\begin{equation}
\label{eq:efmcg}
\sum_{i j} \left[\Braket{i' j'|{\cal C}(f)|i j} - \lambda_f \delta_{i i'}\delta_{j j'}\right] d^{[\rho]k}_{[\mu]i, [\nu]j} = 0\,,\hspace{20pt} f = 2,...,n\,,
\end{equation}
where $\Ket{i j} \equiv \yl{\mu}{i}\yl{\nu}{j}$. Note that the $c_i$ from Eq. \eqref{eq:cgc_singlet} correspond to $d^{\left[1^6\right]1}_{\left[2^3\right]i, [33]6-i}$.
The appropriate eigenvalues $\lambda_f$ can be constructed from the Young tableaux by considering a chain of basis vectors
\begin{equation}
\label{eq:chain}
\yl{\mu}{k} \to \yl{\mu'}{k'} \to \yl{\mu''}{k''} \to ... \,,
\end{equation}
where the $i$-th element is obtained by removing the boxes filled with the $i-1$ largest integers from the corresponding Young tableaux (cf. Eq.\eqref{eq:yamlab}). The eigenvalue $\lambda_f$ is calculated from the diagram with $f$ boxes by summing over all boxes and for each box considering the number of boxes to its left minus the number of boxes above it. For example,
\begin{equation}
\label{eq:ev}
\yng(4,1) \longrightarrow \lambda_f = (0-0) + (1-0) + (2-0) + (3-0) + (0 - 1) = 5\,.
\end{equation}
Solving Eq.~\eqref{eq:efmcg} using Eq.~\eqref{eq:efmcg} and Eq.~\eqref{eq:ev} then gives the Clebsch-Gordan coefficients.

\subsection{Young tableaux, Weyl tableaux and basis vectors}
We now have all the tools required to construct the S-wavefunction. However, for physical applications, we need to map the basis vector $\yl{\mu}{k}$ to a physical state labeled by the basis of the fundamental representation of the corresponding symmetry group. In other words, for an $n$-quark state and a symmetry group $SU(N)$, we need a mapping to the product space of $n$ quarks:
\begin{equation}
\yl{\mu}{n} \to \Ket{\alpha_1, ..., \alpha_n}\,,
\end{equation}
where the $\alpha_i$ are labels for the $N$ basis vectors 
(e.g. $\uparrow$, $\downarrow$ for $SU(2)_s$).
For the \s, living in the singlet of all symmetry groups, this mapping can be straightforwardly constructed. One writes down the corresponding tableaux, fills it with state labels in the unique allowed way (having imposed an ordering on the state labels, e.g. $\uparrow < \downarrow$), symmetrizes over all adjacent numbers in rows and finally antisymmetrizes over all columns. For example, for $SU(3)_F$,
\begin{equation}
\young(14,25,36) \to \young(uu,dd,ss) \to {\cal A}_{123}\Ket{udsuds}\,,
\end{equation}
where ${\cal A}$ denotes the appropriate antisymmetrizer.

While when completed this is the end of the story for the \s, constructing the wavefunction of other (di-)baryons is not quite as straightforward. In the case that a baryon $B$ does not live in a singlet representation of a given $SU(N)$, its wavefunction will carry an additional label denoting the corresponding basis vector in the $SU(N)$ representation. These basis vectors can be labeled by Weyl tableaux. A Weyl tableau has the same shape as the corresponding Young tableau, but is filled with the $SU(N)$ state labels. The combination of Young and Weyl tableaux then uniquely characterizes the state vector, e.g.,
\begin{equation}
\young(12,3) \,\,\young(\uparrow\uparrow,\downarrow)\,.
\end{equation}
In order to construct the appropriate basis vector that has the correct symmetry properties both under quark and $SU(N)$ label exchange, one again resorts to the eigenfunction method. See e.g. \cite{chen_book} for details.


\section{The color-flavor-spin wave function of the \s\ }

\subsection{The Clebsch-Gordon coefficients for the \s}
In case of the \s, we can make use of the special property that it lives in the singlet representation of each of the color-flavor-spin subgroups. This allows us to write each of the basis vectors $\yl{\mu}{k}$ as an appropriate product of the totally antisymmetric symbol of the according group. 

We shall use the following notation. The invariant antisymmetrization symbols of the color, flavor and spin groups will be referred to as
\begin{equation}
\epsilon^c_{\alpha\beta\gamma}\,,\,\,\,\epsilon^f_{ABC}\,,\,\,\,\epsilon^s_{ij}\,,
\end{equation}
respectively. Here, $\alpha,\beta,\gamma,... = r,g,b$ are color indices, $A,B,C,... = u,d,s$ flavor indices and $i,j,k,.. = \uparrow,\downarrow$ spin indices. We will write a vector in the product space as
\begin{equation}
\ket{A_1 \alpha_1 i_1 ,..., A_6 \alpha_6 i_6}\,.
\end{equation}
We further introduce functions $\psi$ of the invariant symbols that generate the different basis functions that transform as singlets under the corresponding $SU(n)$ but have distinct transformation properties under particle exchange. In other words, these label the representations of the symmetric group $S_6$ that correspond to the singlet representations of $SU(n)$. Using Young tableaux, these can easily be shown to be five-dimensional for each of the three different $SU(n)$ groups.
We thus have fifteen different basis functions which we will label through a group index $c,f,s$ and an $S_n$-index $l,m,n$:
\begin{equation}
\psi^{(c,f,s),l}
\end{equation}

In this notation, the $S$-wavefunction takes on the form
\begin{equation}
\label{eq:exp_wf}
\ket{S} = \sum_{l,m,n} d^{lmn} \psi^{c,l}_{\alpha_1,...,\alpha_6}\psi^{f,m}_{A,B,...,F}\psi^{s,n}_{i_1,...,i_6} \ket{A \alpha_1 i_1 ,..., F \alpha_6 i_6}\,,
\end{equation}
where $d^{lmn}$ are the Clebsch-Gordan coefficents.

For the color and flavor $SU(3)$ groups, the basis functions read
\begin{align}
\psi^{(c,f),1}_{a_1,...,a_6} &= \frac{1}{12\sqrt{2}} \left(\epsilon_{a_1a_3a_5}\epsilon_{a_2a_4a_6}+\epsilon_{a_1a_3a_6}\epsilon_{a_2a_4a_5}+\epsilon_{a_2a_3a_5}\epsilon_{a_1a_4a_6}+\epsilon_{a_2a_3a_6}\epsilon_{a_1a_4a_5}\right)\\
\psi^{(c,f),2}_{a_1,...,a_6} &=  \frac{1}{4\sqrt{6}} \left(\epsilon_{a_1a_2a_5}\epsilon_{a_3a_4a_6}+\epsilon_{a_1a_2a_6}\epsilon_{a_3a_4a_5}\right)\\
\psi^{(c,f),3}_{a_1,...,a_6} &=  \frac{1}{4\sqrt{6}} \left(\epsilon_{a_1a_3a_4}\epsilon_{a_2a_5a_6}+\epsilon_{a_2a_3a_4}\epsilon_{a_1a_5a_6}\right)\\
\psi^{(c,f),4}_{a_1,...,a_6} &=  \frac{1}{4\sqrt{6}} \left(\epsilon_{a_1a_2a_4}\epsilon_{a_3a_5a_6}-\frac{1}{3}\epsilon_{a_1a_2a_3}\epsilon_{a_4a_5a_6}\right)\\
\psi^{(c,f),5}_{a_1,...,a_6} &=  \frac{1}{6} \epsilon_{a_1a_2a_3}\epsilon_{a_4a_5a_6}\,,
\end{align}
where $a = A$ $(\alpha)$ for color (flavor).
For the spin $SU(2)$ we have
\begin{align}
	\psi^{(s),1}_{i_1,...,i_6} &= \frac{1}{6} \left(\epsilon_{i_1i_4}\epsilon_{i_2i_5}\epsilon_{i_3i_6}+\epsilon_{i_1i_5}\epsilon_{i_2i_6}\epsilon_{i_3i_4}+\epsilon_{i_1i_6}\epsilon_{i_2i_4}\epsilon_{i_3i_5}\right)\\
	\psi^{(s),2}_{i_1,...,i_6} &= \frac{1}{6\sqrt{2}} \left(\epsilon_{i_1i_3}\epsilon_{i_2i_5}\epsilon_{i_4i_6}+\epsilon_{i_1i_3}\epsilon_{i_2i_6}\epsilon_{i_3i_5}+\epsilon_{i_1i_5}\epsilon_{i_2i_3}\epsilon_{i_4i_6}+\epsilon_{i_1i_6}\epsilon_{i_2i_3}\epsilon_{i_4i_5}\right)\\
	\psi^{(s),3}_{i_1,...,i_6} &= \frac{1}{2\sqrt{6}} \left(\epsilon_{i_1i_3}\epsilon_{i_2i_5}\epsilon_{i_4i_6}+\epsilon_{i_1i_2}\epsilon_{i_3i_6}\epsilon_{i_4i_5}\right)\\
	\psi^{(s),4}_{i_1,...,i_6} &= \frac{1}{2\sqrt{6}} \left(\epsilon_{i_1i_3}\epsilon_{i_2i_4}\epsilon_{i_5i_6}+\epsilon_{i_1i_4}\epsilon_{i_2i_3}\epsilon_{i_5i_6}\right)\\
	\psi^{(s),5}_{i_1,...,i_6} &= \frac{1}{2\sqrt{2}}\epsilon_{i_1i_2}\epsilon_{i_3i_4}\epsilon_{i_5i_6}\,.
\end{align}
These functions are most easily constructed using Young tableaux.

For notational convenience, we display the Clebsch-Gordan coefficients as five $5\times5$-matrices:
\begin{align}
d^{1mn} &= \frac{1}{2\sqrt{10}}\left(\begin{array}{*5{C{1.9em}}}
	0 & 0 & 0 & 0 & 1\\0 & 0 & 0 & 1 & 0\\0 & 0 & 1 & 0 & 0\\ \sqrt{2} & 1 & 0 & 0 & 0\\0 & -\sqrt{2} & 0 & 0 & 0
\end{array}\right),
&d^{2mn} =\frac{1}{2\sqrt{10}}
\left(\begin{array}{*5{C{1.9em}}}
0 & 0 & 0 & 1 & 0\\0 & 0 & 0 & 0 & -1\\-\sqrt{2} & 1 & 0 & 0 & 0\\0 & 0 & -1 & 0 & 0\\0 & 0 & -\sqrt{2} & 0 & 0
\end{array}\right),\nonumber\\
d^{3mn} &=\frac{1}{2\sqrt{10}}\left(\begin{array}{*5{C{1.9em}}}
0 & 0 & -1 & 0 & 0\\\sqrt{2} & -1 & 0 & 0 & 0\\0 & 0 & 0 & 0 & 1\\0 & 0 & 0 & 1 & 0\\0 & 0 & 0 & \sqrt{2} & 0
\end{array}\right),
&d^{4mn} =\frac{1}{2\sqrt{10}}\left(\begin{array}{*5{C{1.9em}}}
\sqrt{2} & 1 & 0 & 0 & 0\\0 & 0 & -1 & 0 & 0\\0 & 0 & 0 & -1 & 0\\0 & 0 & 0 & 0 & 1\\0 & 0 & 0 & 0 & -\sqrt{2}
\end{array}\right),\nonumber\\
d^{5mn} &=\frac{1}{2\sqrt{10}}\left(\begin{array}{*5{C{1.9em}}}
0 & -\sqrt{2} & 0 & 0 & 0\\0 & 0 & -\sqrt{2} & 0 & 0\\0 & 0 & 0 & -\sqrt{2} & 0\\0 & 0 & 0 & 0 & -\sqrt{2}\\0 & 0 & 0 & 0 & 0
\end{array}\right)
\end{align}

\subsection{A compact representation}
It turns out that there exists a very compact representation of the $S$ wavefunction if one utilizes fermionic creation operators. We introduce
$f^\dagger_{\alpha,i}$
as the operator that creates a quark of flavor $f$, color $\alpha$ and spin $i$ and obeys the anticommutation relations
\begin{equation}
\left\{f_{\alpha,i},g^\dagger_{\beta,j}\right\} = \delta_{fg} \delta_{\alpha\beta} \delta_{ij}\,,\hspace{2em} \Big\{f_{\alpha,i},g_{\beta,j}\Big\} = \left\{f^\dagger_{\alpha,i},g^\dagger_{\beta,j}\right\} = 0 \,.
\end{equation}
We can now write Eq.\eqref{eq:exp_wf} as
\begin{equation}
\label{eq:exp_wf2}
\ket{S} = \sum_{l,m,n} d^{lmn} \psi^{c,l}_{A,B,...,F}\psi^{f,m}_{\alpha_1,...,\alpha_6}\psi^{s,n}_{i_1,...,i_6} {A}_{\alpha,i}^{\dagger }{B}_{\beta,j}^{\dagger }{C}_{\mu,k}^{\dagger }{D}_{\nu,l}^{\dagger
}{E}_{\rho,m}^{\dagger }{F}_{\sigma,n}^{\dagger }\ket{\Omega}\,,
\end{equation}
where $\Omega$ is the vacuum state.
This expression can be simplified considerably. By making use of the antisymmetry property of the creation operators and carrying out the summation over flavor indices explicitly, we can put the \s\ wave function into the form
\begin{equation}
\label{eq:comp_wf}
\ket{S} = \frac{1}{N}\left(\epsilon_{\alpha \mu \rho} \epsilon_{\beta \nu \sigma} \epsilon_{im} \epsilon_{jk} \epsilon_{ln} - \epsilon_{\alpha \beta \rho} \epsilon_{\mu \nu \sigma}\epsilon_{im} \epsilon_{jl} \epsilon_{kn}\right) u_{\alpha,i}^{\dagger }u_{\beta,j}^{\dagger }d_{\mu,k}^{\dagger }d_{\nu,l}^{\dagger
}s_{\rho,m}^{\dagger }s_{\sigma,n}^{\dagger }\ket{\Omega}\,.
\end{equation}
Albeit tedious, the derivation is straightforward and we refrain from displaying the steps. The normalization factor $N$ in this representation is non-trivial because in calculating the overlap of initial and final states there are interfering exchange terms.  The calculation of $N$ will be reported elsewhere.  

\section{Hidden color}
\subsection{Decomposition in terms of products of baryons}
\label{sec:hidncolor}
Once the wavefunction of the \s\ has been constructed, we can use it to evaluate various properties of the \s. As a first application, we consider the composition of the \s\ in terms of di-baryon states.

We first note that when a state contains so many quarks that a color singlet can be formed in multiple ways, one cannot naively read off the color content of the wavefunction by mentally inserting the vacuum between color singlet products of operators.  For instance, one might think that Eq.~\eqref{eq:comp_wf} generates a wavefunction which is the product of color singlet baryons of various flavor combinations because in the first term, $u_{\alpha,i}^{\dagger },\,d_{\mu,k}^{\dagger },\,s_{\rho,m}^{\dagger }$ are contracted to a color singlet and similarly for the rest of the creation operators, and likewise for the second term with $u_{\alpha,i}^{\dagger },\,u_{\beta,j}^{\dagger },\,s_{\rho,m}^{\dagger }$ and the rest.  However this is not valid reasoning because the operators anti-commute and can  generate non-color-singlet states as well.  In fact this is what happens.  One must generate the explicit color-flavor-spin wavefunction, either by direct construction as we did, or by acting on the vacuum with the correct creation operator, Eq.~\eqref{eq:comp_wf} or equivalent, which produces the explicit decomposition.

There are a total of eight color-singlet baryon pairs that have the right quark content to contribute to the \s\ wave function: $\Ket{\Lambda\Lambda}$, $\Ket{\Sigma^0 \Sigma^0}$, $\Ket{\Sigma^+ \Sigma^-}$, $\Ket{N \Xi^0}$ and $\Ket{P \Xi^-}$ and their exchanges. From Eq.~\eqref{eq:exp_wf} and the following expressions for the factors, we obtain the overlaps
\begin{multline}
\label{eq:singlet}
\Braket{\Lambda\Lambda|S} = \Braket{\Sigma^0 \Sigma^0|S} = -\Braket{\Sigma^+ \Sigma^-|S} = -\Braket{\Sigma^- \Sigma^+|S} \\= -\Braket{N \Xi^0|S} = -\Braket{\Xi^0 N|S} = \Braket{P \Xi^-|S} = \Braket{\Xi^- P|S} = \frac{1}{\sqrt{40}}\,.
\end{multline}
Products of color singlets hence make up only  $20\%$ of the \s, with the remainder consisting of non-color-singlets.\footnote{Independent evidence of the need for a color octet-octet component in the \s\ comes from Ref.~\cite{MatveevSorbaDeut77}, which explored the possibility that the deuteron might have a small bag-model-like spatially symmetric component.  Those authors comment on the presence of an octet-octet component in a fully anti-symmetrized state of 3 $u$ and 3 $d$ quarks.  The same decomposition -- 20\% singlet-singlet, 80\% octet-octet -- was obtained using operator methods by R. L. Jaffe (unpublished) during his original work on the H-dibaryon (private communication).}  

\subsection{The non-color-singlet components}
\label{sec:88}
Let us investigate the result above, that the color-flavor-spin singlet state of six quarks which is fully anti-symmetric under exchange of any pair of quarks, necessarily includes states in which the individual baryons (here defined to be fully anti-symmetric 3-quark states) are not color singlets. Only by specifying the phases between different superpositions of states, including ones in which the baryons are not color-singlets, can the complete anti-symmetrization be achieved. Our aim in this subsection is to understand the origin of the $\sqrt{4/5}$ coefficient and to deduce the color-flavor-spin properties of those non-color-singlet baryons. 

In Sec.~\ref{sec:Young}, we noted that the color-singlet state $\left[ 2^3 \right]$ and its conjugate flavor-spin singlet $[33]$ are five-dimensional; this is ultimately the origin of the $\sqrt{1/5}$ weight for singlet-singlet baryons.  Clearly, a state composed as in Eq.~\eqref{eq:singlet} is not by itself totally anti-symmetrized between all six quarks in the system, because only the symmetry of the quarks within each baryon, and some collective exchange properties from the baryon flavor-spin specifications, are fixed by the construction.    
The existence of four more terms, each accounting for $\sqrt{1/5}$ of the amplitude in the wave-function, can be seen as follows.  Representing a particular color anti-symmetrization structure by the first tableau below, we see that successive exchanges (3,4),(4,5), (2,3) and (4,5)
\begin{equation}
\label{eq:colorperms}
\young(14,25,36) \to \young(13,25,46) \to \young(13,24,56) \to \young(12,34,56) \to \young(12,35,46)\,,
\end{equation}
produce 4 more distinct realizations of the overall color singlet wave function under permutations of colors within the baryons. Taking the first term to define the basis of color with (123) forming a color singlet gives the eight-term decomposition into products of color-singlet baryons displayed in Eq.~\eqref{eq:singlet}.  We now show that the four other terms in this basis are flavor-singlet products of color-octets.  

As familiar from the construction of the color singlet baryons in the quark model, the representations decompose as {\bf $3 \times 3 \times 3 = 1 + 8_A + 8_S + 10$}, where the subscripts $A$ and $S$ refer to whether the states are anti-symmetric or symmetric in the exchange of the 1-2 entries. When constructing octet baryons, the 3 quarks are in a color singlet -- i.e., totally anti-symmetric in color -- and the flavor-spin states are composed to give a totally symmetric flavor-spin state.  This can be done two ways, producing a spin-3/2 flavor decuplet and a spin-1/2 flavor octet.  In the latter case, the overall symmetric flavor-spin state is produced by taking the sum of the flavor {\bf $8_A$} and {\bf $8_S$} terms, each partnered with the matching antisymmetric or symmetric combination of the 1-2 spins.  Now by analogy, consider the flavor-singlet, color-octet spin-1/2 baryons.  By the same construction as above in flavor-spin, but now in color-spin, we generate states which are totally anti-symmetric within each designated color-octet partition of quarks. But now, with two baryons each of which is a superposition of {\bf $8_A$} and {\bf $8_S$} of color, when the product states are enumerated there are 4 distinct combinations of how the baryon colors can be combined to produce an overall color singlet -- encoded in the last 4 terms of~\eqref{eq:colorperms}. 

\subsection{Impact of color octet components}

To investigate the possible effects of the {\bf 8x8} components of the \s\ wavefunction, we roughly model the $S$ as a bound state of two pointlike tri-quarks each of mass $m_0$, with the (1/5) {\bf 1x1} + (4/5) {\bf 8x8} color structure of Eq.~\eqref{eq:singlet}.  Given the entangled internal wavefunction, Fermi statistics ensures the spatial wavefunction is symmetric and the same for each of the quark constituents.  We base the potential in the {\bf 8x8} channel on the Cornell (heavy quark) potential, devised to account for the splittings among D meson states~\cite{CornellPot75}: 
\begin{equation}
\label{eq:Cornell}
V(r) = \frac{-\alpha }{r}  + \beta\, r ~,
\end{equation}
with $\alpha = 0.52$ and $\beta = 1/(2.34)^2$ (with $\hbar=1$).
 The 20\% color-singlets component of the wave function (the {\bf 1x1} channel) is taken to be non-interacting in this approximation.  A better approximation would take into account the ``back-reaction" associated with confining the {\bf 1x1} components and not treat the tri-quarks as pointlike, but our purpose is only to get an idea of the possible impact of the attraction in hidden-color component of the wavefunction.
 
We take only the Coulomb part of the potential for simplicity.  The 1S bound state in a Coulomb potential has binding energy $\alpha/ (2 r_0)$, where $r_0 = \alpha/\mu $ with $\mu$ the reduced mass. 
In the overall color singlet {\bf 8x8} channel, the potential is 9/4 times stronger than in the $\mathbf{3x\bar{3}}$ channel.\footnote{This is just the ratio of quadratic Casimirs for the octet and triplet representations.  The quadratic Casimir $C_2$ is defined by $T^a T^a = C_2\, \mathbb{I}$ where $T^a$ and $\mathbb{I}$ are respectively the generators and unit matrix in the representation of interest; for the fundamental and adjoint representations of SU(N), $C_F = (N^2 - 1)/ (2N)$ and $C_A = N$. }  Thus the binding energy in the {\bf 8x8} channel is $(m_0 /4) \,(9\, \alpha/4)^2$.   Since only 4/5 of the state feels the potential, the overall binding energy is 4/5 of this Coulomb value, or $(81/80)m_0 \alpha^2$.  With $\alpha = 0.52$ from the Cornell potential, this gives a bound state mass 14\% lower than the sum of the constituent masses.  

Taking $m_0 = m_\Lambda$ gives a binding energy of 305 MeV or $m_S = 1925$ MeV.  This is just a rough estimate to indicate the potential impact of excluding the octet-octet channel from mass estimates.  A more realistic treatment with fuzzy rather than point-like constituents would likely give less deep binding.  It is worth noting that $m_S$ up to 2050 MeV is low enough that the \s\ lifetime is much greater than the age of the Universe and compatible with the \s\ being a significant component of the Dark Matter~\cite{xf23a,fw23a} and is allowed by laboratory experiments~\cite{fS22,fw23a}.
Another noteworthy result of this model is that the characteristic size of the system, $r_0 = (2/ m_0) /(9/4 \alpha) = 0.3 \,(m_\Lambda/m_0) $ fm is much smaller than for known mesons and baryons made of light quarks.

\subsection{Alternate ways to represent the state}

The color-flavor-spin wavefunction of the \s\ explicitly given in Eq.~\eqref{eq:exp_wf} and associated expressions is the clearest way to characterize the color-flavor-spin structure of the \s.  Equivalent, but less explicit in terms of the color-flavor-spin content, are representations in terms of products of creation operators acting on the vacuum, as in Eq.~\eqref{eq:comp_wf}.  As we have shown, the overall-singlet 6-quark state includes components which do not factor into color-singlet baryons.  Such ``hidden color" within a state cannot be determined by simply inserting the vacuum state between color-singlet products of three quark creation operators appearing in the product of six creation operators acting on the vacuum; rather one must resolve the state into a basis of color-flavor-spin eigenstates, c.f., Eq.~\eqref{eq:exp_wf} et seq.

Concrete examples of the non-uniqueness of the operator representation of the wavefunction are given in Ref.~\cite{DGH86}, which discusses several different ways to write a color-flavor-spin singlet H-dibaryon in terms of sums of products of quark creation operators acting on the vacuum. Equation~\eqref{eq:comp_wf} is a still more compact representation than those presented in Ref.~\cite{DGH86} and, importantly, we have demonstrated that it creates a color-flavor-spin singlet.  The principle representation of~\cite{DGH86} is a sum of six terms, each of which is a product of color-singlet three-quark operators, with no explicitly ``crossed" color indices as in the second term of ~\eqref{eq:comp_wf}.  However even though the color contractions in each term of the principle representation of Ref.~\cite{DGH86} factor into color singlet 3-quark operators, like the first term of Eq.~\eqref{eq:comp_wf}, all the corresponding explicit wavefunctions are necessarily equivalent to Eq.~\eqref{eq:exp_wf} and have amplitude $\sqrt{4/5}$ to be a product of color octet baryons.\footnote{The aim of~\cite{DGH86} was calculating the lifetime of the H-dibaryon, which they found to be $\approx 10^{-8}$ s. It should be noted that the calculation of~\cite{DGH86} does not apply to a compact sexaquark due to the spatial overlap of the initial and final states being entirely different for the H and the \s.  Ref.~\cite{DGH86} assumed that the spatial wavefunction of the quarks in the H-dibaryon is similar to or more extended than the wavefunction of individual baryons, but this is not the case for the \s~\cite{fzNuc03,fS17,fS22}.  Under the deep-binding hypothesis the \s\ is much more compact than an isolated baryon, as found for instance in the Cornell potential model presented above as well as in an empirical model of hadron sizes based on the Compton wavelengths of the state and the lightest meson to which it couples~\cite{fS17}.  As a result, the quantum-mechanical amplitude \gtilde, for all six quarks in two separate baryons to fluctuate to the compact spatially-symmetric configuration of the \s, is very small due to the hard-core repulsion of individual color-singlet baryons~\cite{fzNuc03}. Using a realistic, small \gtilde\ implies an \s\ di-baryon lifetime greater than the age of the Universe~\cite{fzNuc03,fS22};  see~\cite{fw23a} for a survey of experimental constraints on \gtilde.}

\section{Summary}
We have derived an explicit expression for the color-flavor-spin wavefunction of a spatially symmetric $uuddss$ state, consistent with both Fermi statistics and the desired color-flavor-spin-singlet quantum numbers of the \s\ or H-dibaryon.  We also present a compact creation operator which creates the desired overall-singlet state.   

The explicit wavefunction shows that the overall-singlet state has an amplitude of only $1/\sqrt{5}$  to be a product of color-singlet baryons.   
We used a simple potential model -- the Coulombic part of the Cornell potential -- to estimate the impact on the binding energy of including the octet-octet component, and found inclusion of this component to considerably reduce the mass of the state.  

Lattice QCD calculations of the mass of the lightest state in the $uuddss$ channel often create or probe the state with a product of color-singlet baryon operators with total isospin 0 and strangeness -2.   Since the state is generated by fermionic operators, it is automatically fully-antisymmetric under interchange of any pair of quarks.  However without keeping both terms in Eq.~\eqref{eq:comp_wf} or the equivalent, the state is not an overall-singlet in flavor even though it has the required quark content.  Using the Clebsch-Gordon tables of Ref.~\cite{SU3Clebsch64}, one can see that the product of two flavor octets has an  amplitude of only $1/\sqrt{8}$ to be a flavor singlet, and amplitudes $\sqrt{27/40}$ and $ \sqrt{1/5}$ to be a flavor {\bf 27} or {\bf 8}, respectively. In the L\"uscher method, this delays the emergence of the true ground state.  Furthermore, the color magnetic (hyperfine) interaction increases the masses of the higher representations relative to the overall singlet state~\cite{jaffe:H}, so unless the individual contributions are resolved, the inferred mass will be larger than the true one.\footnote{With the H-dibaryon apparently being a $\Lambda \Lambda$ molecule or near-threshold resonance, its spatial wavefunction is like a deuteron's:  two well-separated 3-quark baryons, not spatially symmetric at all.  So its relative contribution to a plateau depends in general on the spatial structure of the creation and probing operators.}

Since a major hurdle to robust hadron mass determination using lattice QCD for physical quark masses is the growth of noise with Euclidean time, especially for multi-light-quark states~\cite{LepageTASI}, employing the operator derived here should facilitate discovering or definitively excluding a possible deeply bound state.

\section*{Acknowledgements}
GRF thanks Zihui Wang for helpful discussions.  The research of GRF has been supported by National Science Foundation Grant No.~PHY-2013199 and by the Simons Foundation.  NW thanks the Niels Bohr Institute for its support during the completion of this work.


\begin{thebibliography}{99}

\bibitem{ALICE_LamLam19}
{\bf ALICE} Collaboration, S.~Acharya {\em et~al.}, {\it {Study of the
  $\Lambda$-$\Lambda$ interaction with femtoscopy correlations in pp and p-Pb
  collisions at the LHC}},  {\em Phys. Lett. B} {\bf 797} (2019) 134822,
  [\href{http://arxiv.org/abs/1905.07209}{{\tt arXiv:1905.07209}}].

\bibitem{SasakiHALQCD20}
{\bf HAL QCD} Collaboration, K.~Sasaki {\em et~al.}, {\it {$\Lambda\Lambda$ and
  N$\Xi$ interactions from lattice QCD near the physical point}},  {\em Nucl.
  Phys. A} {\bf 998} (2020) 121737,
  [\href{http://arxiv.org/abs/1912.08630}{{\tt arXiv:1912.08630}}].

\bibitem{greenMainzH21}
J.~R.~Green, A.~D.~Hanlon, P.~M.~Junnarkar and H.~Wittig,
Phys. Rev. Lett. \textbf{127}, no.24, 242003 (2021)
doi:10.1103/PhysRevLett.127.242003
[arXiv:2103.01054 [hep-lat]].

\bibitem{jaffe:H}
R.~Jaffe, {\it Perhaps a stable dihyperon...},  {\em Phys. Rev. Lett.} {\bf 38}
  (1977) 195,
  [\href{http://arxiv.org/abs/http://arXiv.org/abs/nucl-th/9912031}{{\tt
  http://arXiv.org/abs/nucl-th/9912031}}].

\bibitem{fzNuc03}
G.~R.~Farrar and G.~Zaharijas,
Phys. Rev. D \textbf{70}, 014008 (2004)
doi:10.1103/PhysRevD.70.014008
[arXiv:hep-ph/0308137 [hep-ph]].

\bibitem{fS17}
G.~R.~Farrar,
[arXiv:1708.08951 [hep-ph]].

\bibitem{fudsDM18}
G.~R.~Farrar,
[arXiv:1805.03723 [hep-ph]].
\bibitem{fS22}
G.~R. Farrar, {\it {A Stable Sexaquark: Overview and Discovery Strategies}},
  \href{http://arxiv.org/abs/2201.01334}{{\tt arXiv:2201.01334}}.

\bibitem{xf23a}
X.~Xu and G.~R.~Farrar,
Phys. Rev. D \textbf{107}, no.9, 095028 (2023)
doi:10.1103/PhysRevD.107.095028
[arXiv:2101.00142 [hep-ph]].

\bibitem{fw23a}
G.~R.~Farrar and Z.~Wang,
[arXiv:2306.03123 [hep-ph]].

\bibitem{LepageTASI}
G.~P.~Lepage,
[arXiv:hep-ph/0506330 [hep-ph]].

\bibitem{pplSwfn16}
W.~Park, A.~Park and S.~H.~Lee,
Phys. Rev. D \textbf{93}, no.7, 074007 (2016)
doi:10.1103/PhysRevD.93.074007
[arXiv:1602.05017 [hep-ph]].
\bibitem{rosnerH86}
J.~L.~Rosner,
Phys. Rev. D \textbf{33}, 2043 (1986)
doi:10.1103/PhysRevD.33.2043


\bibitem{MackenzieThacker85}
P.~B.~Mackenzie and H.~B.~Thacker,
Phys. Rev. Lett. \textbf{55}, 2539 (1985)
doi:10.1103/PhysRevLett.55.2539



  \bibitem{hammermesh}
M.~Hammermesh,{\it Group Theory and its applications to physical problems}, Addison-Wesley, 1962.

\bibitem{chen_book}
Chen, J. Q., Ping, J. \& Wang, F. (1989). \emph{Group representation theory for physicists (Vol. 7)}. Singapore: World Scientific.

\bibitem{chen_review}
Chen, J. Q., Gao, M. J. \& Ma, G. Q. (1985). \emph{The representation group and its application to space groups}. Reviews of modern physics, 57(1), 211.

\bibitem{CornellPot75}
E.~Eichten, K.~Gottfried, T.~Kinoshita, J.~B.~Kogut, K.~D.~Lane and T.~M.~Yan,
Phys. Rev. Lett. \textbf{34}, 369-372 (1975)
[erratum: Phys. Rev. Lett. \textbf{36}, 1276 (1976)]
doi:10.1103/PhysRevLett.34.369

\bibitem{DGH86}
J.~Donoghue, E.~Golowich and B.~R.~Holstein,
Phys. Rev. D \textbf{34}, 3434 (1986)
doi:10.1103/PhysRevD.34.3434

\bibitem{MatveevSorbaDeut77}
V.~A.~Matveev and P.~Sorba,
Lett. Nuovo Cim. \textbf{20}, 435 (1977)
doi:10.1007/BF02790723 and 
Nuovo Cim. A \textbf{45}, 257 (1978)
doi:10.1007/BF02724667




\bibitem{SU3Clebsch64}
P.~S.~J.~McNamee and F.~Chilton,
Rev. Mod. Phys. \textbf{36}, 1005-1024 (1964)
doi:10.1103/RevModPhys.36.1005





\end{thebibliography}

\end{document}